\documentclass[a4paper,11pt]{article}
\pdfoutput=1 

\usepackage{epsfig}
\usepackage{graphicx}                  
\usepackage{ae}
\usepackage{amsmath}
\usepackage{amssymb}
\usepackage{graphics}
\usepackage{dcolumn}
\usepackage{bm}
\usepackage{lineno}
\usepackage{caption}
\usepackage{ragged2e}
\usepackage{subcaption}

\renewcommand{\thefootnote}{\fnsymbol{footnote}}
\title{\bf Charging-up and reverse charging-up phenomena in a double-mask triple GEM detector}
\date{}

\begin{document}

\maketitle
	\flushbottom
\vspace*{-1cm}
\centering

\author{\bf S.~Mandal$^{1}$,}
\author{\bf S.~Ghosh$^{2}$,}
\author{\bf S.~Das$^{1}$,}
\author{\bf S. Biswas$^{1}$$^*$}
\let\thefootnote\relax\footnotetext{$^*$Corresponding author. 

\hspace*{0.4cm}E-mail: saikat@jcbose.ac.in, saikat.ino@gmail.com }

\vspace*{0.5cm}

	$^1${{Department of Physical Sciences, Bose Institute, EN-80, Sector V, Kolkata-700091, India}
	
	$^2${{Department of Physics, Presidency University, Kolkata, India}

\vspace*{0.5cm}
\centering{\bf Abstract}
\justify

The Gas Electron Multiplier (GEM) detectors are widely used in high-energy physics (HEP) experiments as tracking devices because of their excellent position resolution and to handle high particle rates capability. Charging-up effect is a well known phenomenon in GEM detectors because of the presence of the dielectric medium - Kapton in the foil. Charging-up of GEM foil takes place when it is exposed to high radiation after application of high voltage. A new phenomenon of reverse charging-up, a complementary behaviour is also observed when the irradiation rate is reduced, where the gain relaxes gradually towards its initial value. In this study, the charging-up and reverse charging-up effects are investigated for a double-mask triple GEM chamber operated with an Argon and Carbon dioxide (70/30) gas mixture. The measurements provide a detailed understanding of the gain variation under irradiation and its stabilisation behaviour. The experimental setup, methodology and results are presented in this article.

\vspace*{0.25cm}


\section{Introduction} \label{intro}
The Gas Electron Multiplier (GEM) detectors, first introduced by Fabio Sauli in 1997~\cite{Sauli}, have become an important class of micro-pattern gaseous detectors (MPGD) for high energy physics (HEP) experiments. Their ability to provide high rate capability, good position resolution, and low ion back flow makes them suitable for tracking and triggering applications in modern HEP experiments~\cite{Buzulutskov,Ketzer,Sharma2006}. In recent time, large experiments such as ALICE at the LHC \cite{ALICE_TDR} have accepted triple GEM technology for the TPC because of its TPC for handling high interaction rate and low ion back flow, CMS are using triple GEM chambers for the upgraded muon detection system as the tracking systems \cite{CMS}. Future experiment CBM at FAIR is also considering the GEM chambers as the tracking detectors due to the foreseen high interaction rates \cite{Adak2016,Chatterjee2021,cbm,Galatyuk,ALICE2021}.

A GEM foil is typically a thin (50~$\mu$m) Kapton layer cladded with 5~$\mu$m thick copper on both sides, perforated with a dense pattern of microscopic holes created using photolithographic technique~\cite{Bachmann2002,RD51}. Depending on the fabrication method, foils are classified as either single-mask (SM) or double-mask (DM) types. While single-mask technology is well suited for large area chambers, the double-mask process produces more symmetric bi-conical holes, which directly influence the uniformity of the electric field and the stability of the gain~\cite{uniformity_1,mandal_2024,Abbaneo2003}. 

At the start of operation, after the application of high voltage (HV), the performance of such GEM detectors is influenced by the dielectric medium within the active volume, which alters the detector response when exposed to radiation. This effect, known as the charging-up phenomenon, leads to a time-dependent increase in the gain.

In GEM operation, the charging-up phenomenon arises in the dielectric medium within the holes. When exposed to radiation, the accumulation of charges on the Kapton surface modifies the effective electric field, leading to a time-dependent variation of the gain~\cite{Adak2016, Bachmann2002, Bressan2000, s_chatterjee_charging_up_1}. At the beginning of operation, this effect is often observed as an increase in the gain until a steady state is reached. The phenomenon is equivalent to the charging of an RC circuit and is studied in detail for both single-mask and double-mask GEM prototypes \cite{s_chatterjee_charging_up_2, chatterjee_2023_charge, Altunbas2002, abbrescia2008, Silva2019}. 

Several studies have shown that hole geometry plays a crucial role in the charging-up behaviour and long-term stability of GEM detectors~\cite{Bressan2000,s_chatterjee_charging_up_1,s_chatterjee_charging_up_2,chatterjee_2023_charge}.

The present work focuses on a double-mask triple GEM detector prototype operated with an Ar/CO$_2$ gas mixture in 70/30 volume ratio. The aim is to study the behaviour of gain and energy resolution under continuous irradiation. These studies provide crucial role into the suitability of GEM technology for long-term high rate operation.

\section{Experimental Setup}\label{ES}

The prototype used in this study is a 10~cm~$\times$~10~cm triple GEM detector fabricated at CERN with standard stretched double-mask GEM foils \cite{Sharma2006, RD51,DelPapa2018}. The drift gap, two transfer gaps, and the induction gap are maintained at 3 mm, 2 mm, 2 mm, and 2 mm, respectively.  

The chamber is biased using a resistive voltage divider chain connected to a single negative HV channel. 10~M$\Omega$ protection resistors are connected to the drift plane and top of all the three GEM foils. 560~k$\Omega$ resistors are connected between the top and bottom of all three GEM foils. 1~M$\Omega$ resistors are connected between the drift and top of the first GEM foil, between the GEM foils and between bottom of the thirst GEM foil. The schematic of the HV distribution chain for the triple GEM prototype is shown in Ref.~\cite{s_chatterjee_charging_up_1}. At an operating voltage of -~4050 V, the corresponding electric fields in the drift, transfer, and induction regions are about 2.21~kV/cm, 3.31~kV/cm, and~3.31 kV/cm, respectively, while the voltage ($\Delta$V) across each GEM foil is $\sim$~370~V~\cite{Bucciantonio, Sahu2017}. 

\begin{figure}[htb!]
\centering
\includegraphics[scale=0.3]{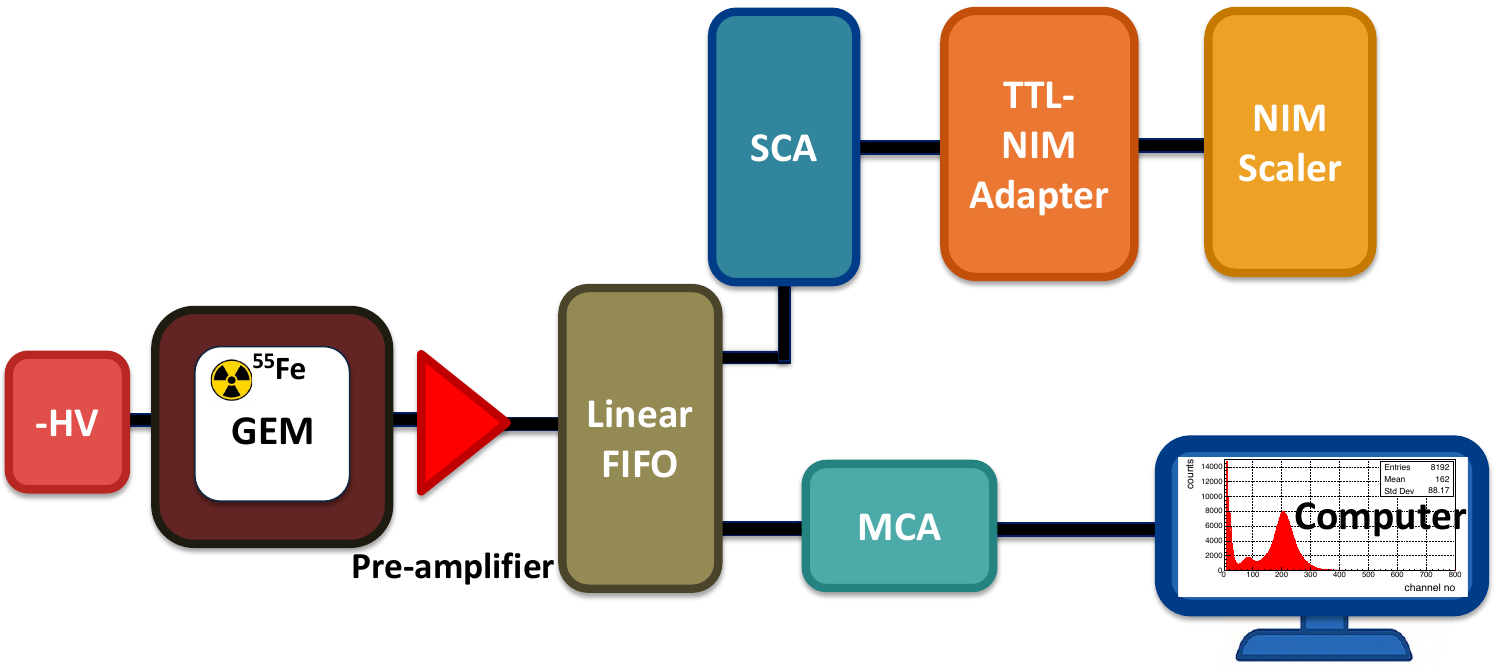}
\caption{(Colour online) Schematic representation of the electronic setup~\cite{s_chatterjee_charging_up_1, mandal_2026}.}
\label{fig_2}
\end{figure}

As the exposure, an $^{55}$Fe X-ray source (energy: 5.9~keV) is collimated using a G-10 made collimator irradiating an area of $50\,\mathrm{mm}^2$ of the prototype. The chamber readout consists of 9 pads, each having dimensions of $9\,\mathrm{mm} \times 9\,\mathrm{mm}$ at the central part of the prototype. For this work, the signals are summed using CERN-provided sum-up boards and processed through a low-noise charge-sensitive preamplifier (VV50-2)~\cite{preamplifier}. The gain and a shaping time of the preamplifier are 2~mV/fC and 300~ns respectively. The output from the preamplifier is sent to a linear Fan-In Fan-Out (FIFO) module to make multiple replica of the same analog signal. One output of the FIFO is fed to a Multi-Channel Analyser (MCA) to record the X-ray spectra on a desktop computer. Another output from the FIFO is connected to a Single-Channel Analyser (SCA), whose output is passed through a TTL-to-NIM adapter, and the signals above the noise threshold are counted using a NIM scaler. The final count rate is recorded from the NIM scaler. The SCA threshold is kept constant at 0.4~V throughout the measurements~\cite{mandal_2024}. The schematic of the electronic circuit for the DM triple GEM prototype is shown in Figure \ref{fig_2}, which is same as shown in Ref~\cite{s_chatterjee_charging_up_1} and Ref~\cite{mandal_2026}. The chamber is operated with a pre-mixed Ar/CO$_2$ (70/30) gas mixture at a constant flow rate of about 3.4-4 l/h, maintained using a Vögtlin gas flow meter. Ambient parameters such as temperature (t), pressure (p), and relative humidity (RH) are continuously recorded using an in-house developed data logger \cite{Zhao2014, chatterjee_2023_rh, Anderson2020, mandal_datalogger}. With this setup, specially the time dependence of the gain and energy resolution during a transition between the low-rate and high-rate of X-ray irradiation is investigated.

\section{Observables}\label{obs}

In this study, the gain, energy resolution, count rate, divider current, ambient temperature ($t$), pressure ($p$) and relative humidity (RH) are measured as a function of time~\cite{Chatterjee2021,mandal_2024}. The typical spectrum obtained at an applied HV of -~4050~V corresponding to a $\Delta{V}$~=~370~V accross each GEM foil is displayed in Figure~\ref{fig11}. The spectrum shows a clear main peak at 5.9~keV, a 2.9~keV escape peak along with the noise peak. The 5.9~keV main peak is fitted with a Gaussian function to calculated the gain and energy resolution of the chamber. The total output charge for further analysis is estimated by the mean value of this Gaussian fit, preamplifier gain, and MCA calibration factor.

\begin{figure}[htb!]
	\begin{center}
		\includegraphics[scale=0.45]{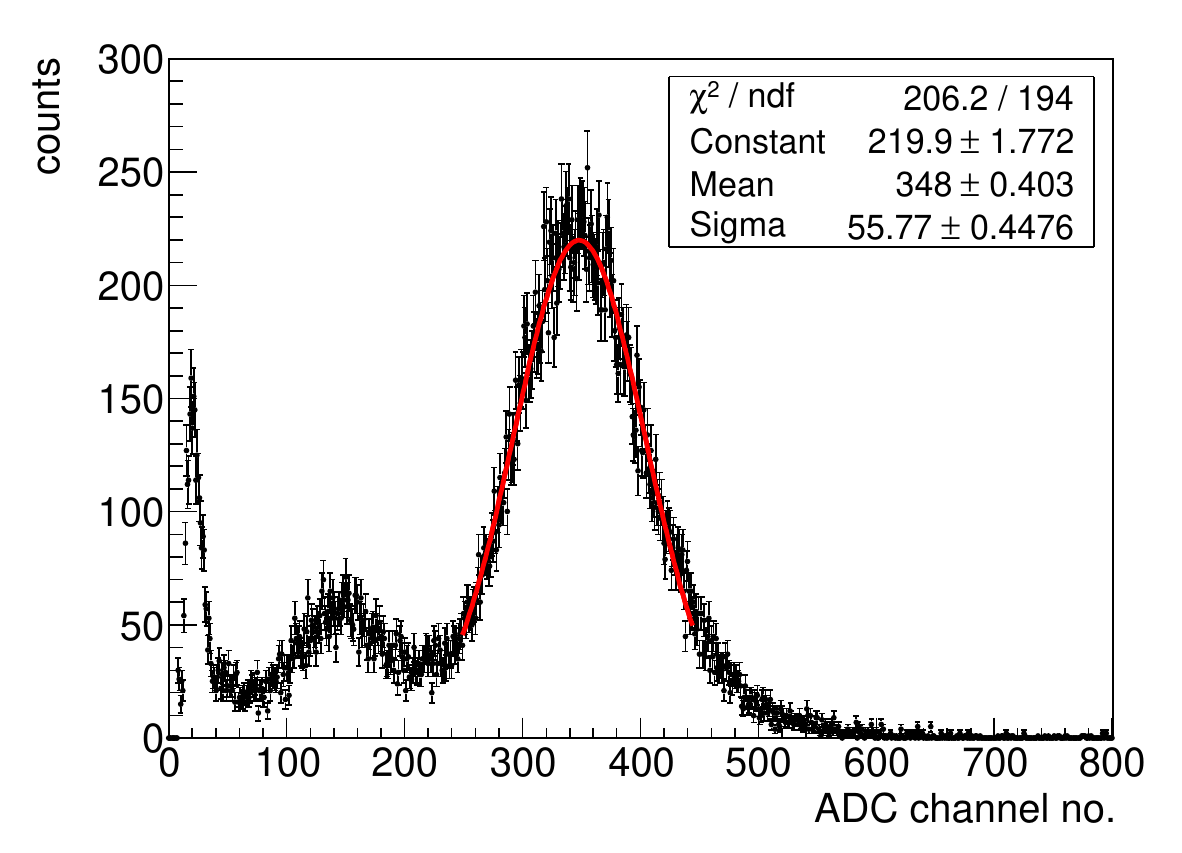}
		\caption{(Colour online) Typical $^{55}$Fe spectrum in Ar/CO$_2$ (70/30 volume ratio) gas mixture at -~4050~V. The $\Delta V$ across each of the GEM foils is $\sim$~370~V.}\label{fig11}
	\end{center}
\end{figure}

The gain is defined as,  

\begin{equation}
\begin{aligned}
gain &= \frac{output \; charge}{input \; charge} \\
&= \frac{\frac{Pulse \; height}{2 \; \text{mV}} \; fC}{n_{primary} \times e}
\end{aligned}
\end{equation}

where $n_{primary}$ (212 for Ar/CO$_2$ 70/30 ratio) is the number of primary electrons and $e$ is the electronic charge.  

The pulse height is obtained using the MCA calibration factor. The MCA is calibrated with a pulse with known height from a function generator, and the relation between the pulse height and the mean MCA channel number is expressed as,  

\begin{equation}
Pulse \; height \; (V) = MCA \; Channel \; no. \times 0.0014 \; + \; 0.14
\end{equation}

The full width at half maximum (FWHM) of the Gaussian fitted spectra is used to define the energy resolution of the chamber. The energy resolution is calculated using the relation,  

\begin{equation}
Energy \; resolution = \frac{\sigma \times 2.355}{\mu}
\end{equation}

where, the mean $\mu$ and $\sigma$ are obtained from the Gaussian fitting of the energy spectra. The details are described in Ref.~\cite{mandal_2024,Chatterjee_thesis}.

As mentioned earlier, the main goal is the study of gain and energy resolution stability of triple GEM chamber under continuous irradiation. In this article, during continuous operation, both the charging-up and reverse charging-up behaviour of the double mask GEM detector are observed when the rate of X-ray is changed. When external radiation is exposed to the GEM chamber after applying the voltage, Kapton - the dielectric medium  present in the active volume of the detector, changes the behaviour of the chamber. As a result, the gain of the detector increases initially and then reaches a plateau asymptotically. This increase in gain is due to the well-known charging-up phenomenon of the dielectric medium. During stability study of gaseous detectors, usually the instantaneous gain is measured with a defined time interval, the temperature (T) and pressure (p) are also monitored for each point, the gain and T/p correlation is fitted with some function and finally the instantaneous normalised gain is plotted as a function of time. The constant value of normalised gain within some fluctuation limit with time tells that the detector works stably. During this whole study when the high rate of X-ray is introduced an  increase in the gain is observed, when the rate is reduced in presence of applied high voltage again an increase in the gain is observed from its present value. The gain variation in these phases follows an exponential trend analogous to an RC circuit response~\cite{Bressan2000,s_chatterjee_charging_up_1}.

\begin{figure*}[htb!]
\centering
\includegraphics[scale=0.64]{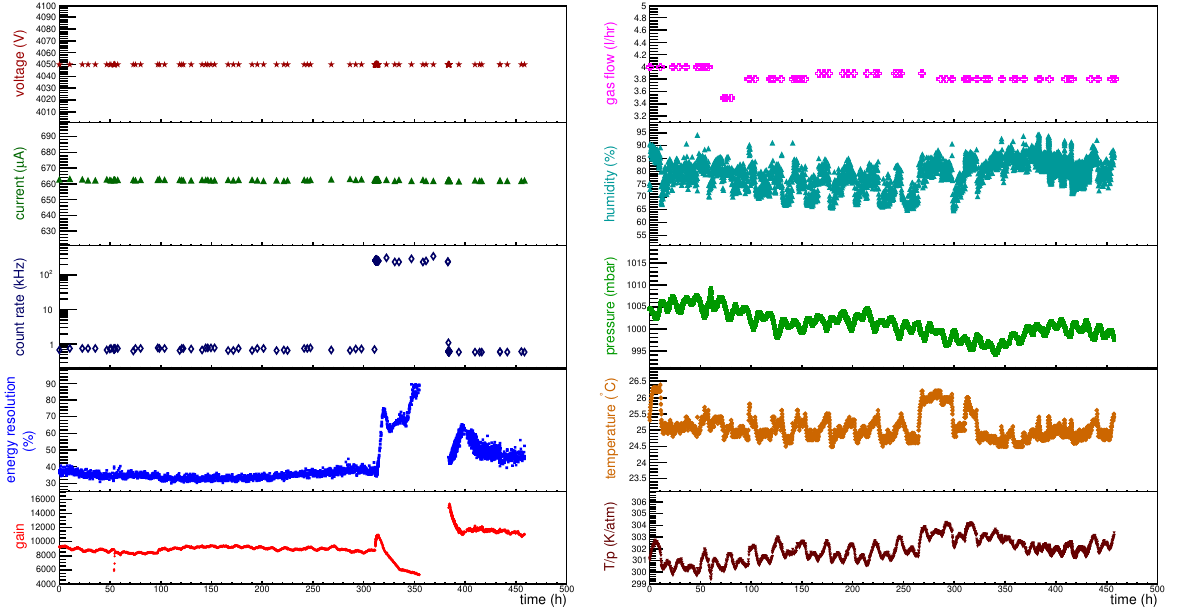}
\caption{(Colour online) Gain, energy resolution, $T/p$, volt, gas-flow rate, count and current as a function of time.}
\label{fig2}
\end{figure*}

\section{Results}\label{res}

In this study the X-ray energy spectra for 1 minute are stored in 10 minutes interval. The ambient temperature (t), pressure (p) and relative humidity (RH) are also measures with the same 10 minutes interval. The value of the gas flow rate, applied voltage, bias current through the voltage divider circuit, count rate are also measured time to time manually. The gain and energy resolution are calculated fitting the main peak of the $^{55}$Fe X-ray spectrum by Gaussian function as explained in Sec~\ref{obs}. All the measured observables and the calculated values as a function of time are shown in Figure~\ref{fig2}. 

\begin{figure*}[htb!]
\centering
\includegraphics[scale=0.65]{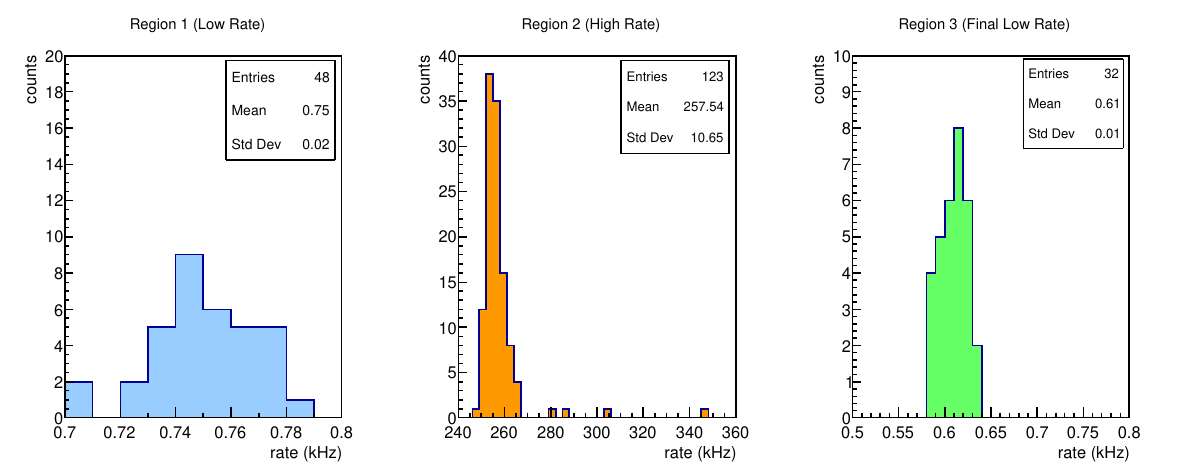}
\caption{(Colour online) Histogram distributions of count rates to determine average values during low, high, and second low-rate phases.}
\label{fig3}
\end{figure*}

The distributions of the count rate for three time zones are shown in Figure~\ref{fig3}. It is observed from Figure~\ref{fig3} that the count rate distributions clearly separates the whole measurements into three phases: initial low, high, and final low rate regions. During the high-rate phase the count rate increases significantly, reflecting the higher particle flux. The clear difference in the mean position of the peaks of the distributions confirms that the source exposure and the stability of the detector response in different irradiation regimes. These distributions are important to establish the operating range and to quantify the transition behaviour 
between low and high-rate conditions.

Initially the average count rate is found to be ~0.75~kHz. It is observed from Figure~\ref{fig2} that the detector shows stable gain and energy resolution during the initial low rate phase, with only small fluctuations that remain within acceptable limits for long-term operation. A sudden change in gain is noticed at about 50 hours due to the replacement of the gas cylinder, which causes a temporary variation in the measured parameters. When the irradiation rate increases, the gain first rises sharply as a result of increased charge accumulation inside the GEM holes i.e. in the Kapton because of the charging-up, but then gradually decreases as the charge build-up modifies the local electric field and reduces the effective multiplication. At the same time, the energy resolution became worst due to increased statistical fluctuations in avalanche size and distortions in the electric field geometry. When the irradiation rate decreases back to a low value, the gain increases again and eventually stabilises close to the initial values, reflecting the redistribution and leakage of stored charges. The energy resolution improved as the fields recover towards their initial configuration, except during a short interval between 311 and 314 hours where both gain and resolution increase simultaneously, likely due to a temporary variation in the local field charging up.

\begin{figure*}[htb!]
\centering
\includegraphics[scale=0.65]{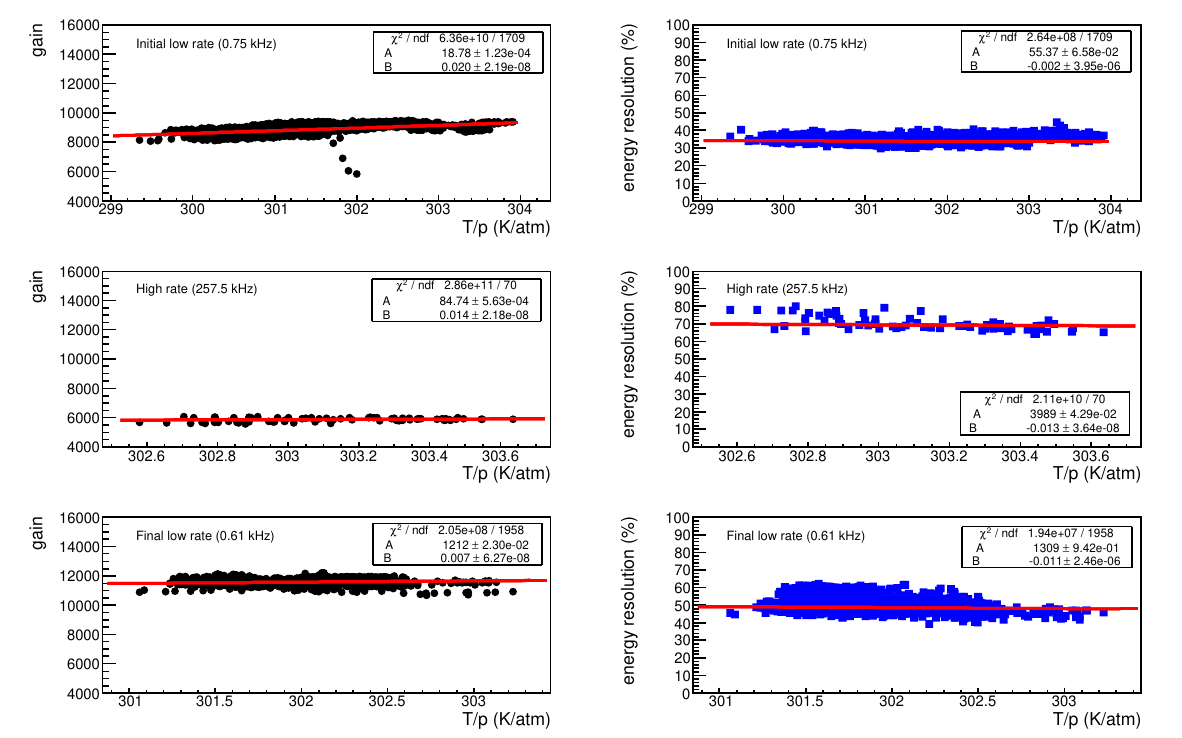}
\caption{(Colour online) Correlation of gain and energy resolution with $T/p$ for the initial low rate, intermediate high rate and final low rate regions.}
\label{fig4}
\end{figure*}

The gain and energy resolution as functions of T/p (T=t+273) for the initial and final low rate region and intermediate high rate region are plotted in Figure~\ref{fig4}. The gain and energy resolutions vs. T/p plots are fitted by 
\begin{equation}
\centering
G (T/p) = A e^{(B \frac{T}{p})}\label{correlation}
\end{equation} 
and
\begin{equation}
 \centering
 energy \; resolution~(T/p) = A' e^{(B' \frac{T}{p})}\label{correlation_er}  
 \end{equation}
respectively. Where, $A$, $B$ and $A'$, $B'$ are fit parameters. It is observed from Figure~\ref{fig4} that both the gain and energy resolution show clear dependence on the $T/p$ ratio for all three irradiation phases. The fitted exponential functions show the expected correlation between detector performance and ambient parameters. As $T/p$ increases, the gas density decreases, leading to higher mean free paths for electron-ion collisions and therefore higher gas amplification. This explains the observed rise in gain with $T/p$. Conversely, energy resolution value decreases at higher $T/p$ giving a better resolution. The consistency of this trend across low and high irradiation phases highlights the importance of correcting measured gain and energy resolution for $T/p$ variations in order to separate true charging-up effects from ambient conditions.

\begin{table}[htb!]
\centering
\resizebox{0.45\textwidth}{!}{
\begin{tabular}{|l|c|c|c|c|}
\hline
Region & A & B & A' & B' \\
\hline
Initial Low Rate (0.75 kHz) & 18.78 & 0.02042 & 55.37 & -0.002 \\
\hline
High Rate (257.5 kHz)       & 84.74 & 0.01398 & 3989 & -0.013 \\
\hline
Final Low Rate (0.61 kHz)   & 1212 & 0.007472 & 1309 & -0.011 \\
\hline
\end{tabular}
}
\caption{Fit parameters for gain (A, B) and energy resolution (A', B')}
\label{tab1}
\end{table}

\begin{figure*}[htb!]
\centering
\includegraphics[scale=0.65]{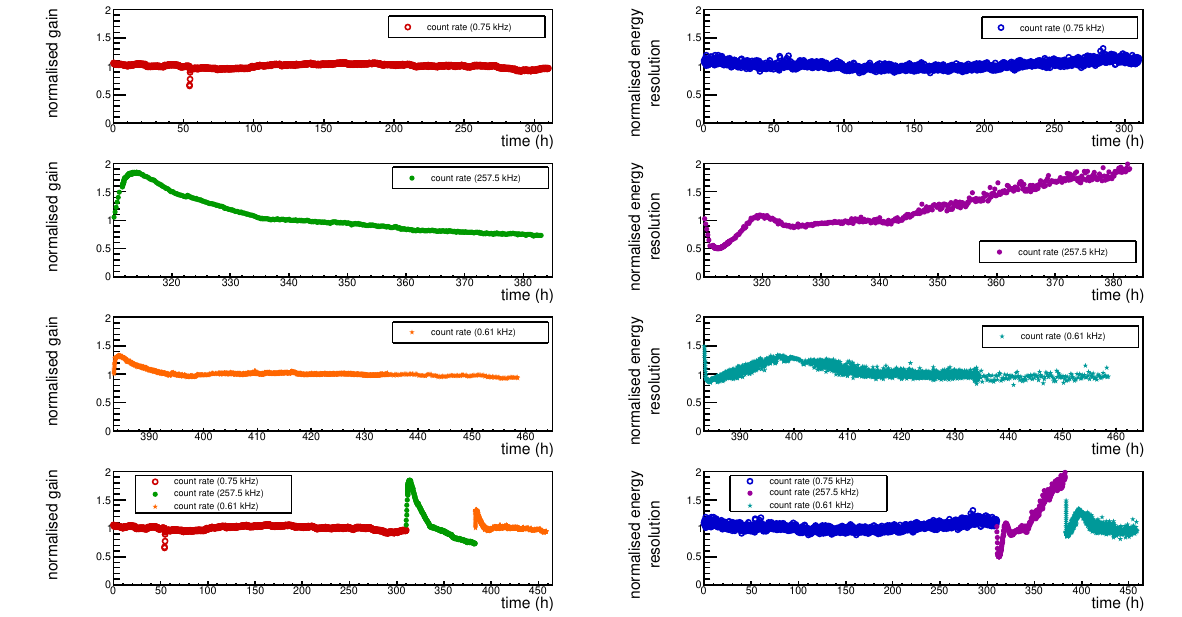}
\caption{(Colour online) Normalised gain and normalised energy resolution as a function of time for the entire period and also for the three different rate phases.}
\label{fig5}
\end{figure*}

After fitting the measured gain and energy resolution by the functions~\ref{correlation} and \ref{correlation_er} respectively, the parameters $A$, $B$ and $A'$, $B'$ are extracted for all three regions and the values are  tabulated in Table~\ref{tab1}. The measured gain and energy resolution are then normalised by the fit functions ~\ref{correlation} and \ref{correlation_er} and using the parameters $A$, $B$ and $A'$, $B'$. The normalised gain and normalised energy resolution as function of time are shown in Figure~\ref{fig5}. It is observed from Figure~\ref{fig5} that the normalised gain and energy resolution remain stable during the initial low rate phase, confirming that the detector is well-behaved under low radiation exposure and the normalised gain is found to be nearly 1. When the exposure rate is changed from 0.7k~kHz to 257~kHz, the normalised gain initially increases from its present value of $\sim$~1 to nearly 2, as because of charging-up effect the charge accumulation increases the local field strength, but subsequently decreases once surface charging saturates and modifies the field geometry. The normalised energy resolution, in contrast, changes inversely. During the final high-rate to low-rate transition, the normalised gain again rises briefly as trapped charges begin to relax, but then decays gradually to stable values, closely doing an RC discharge process. A similar recovery trend is observed for the normalised resolution. The phenomenon of increase in normalised gain during the high rate to low rate transition is defined as the reverse charging-up. These results shows that the gain and resolution dynamics are strongly coupled to the charging-up and reverse charging-up of the GEM dielectric surfaces.

\begin{figure*}[htb!]
\centering
\includegraphics[scale=0.65]{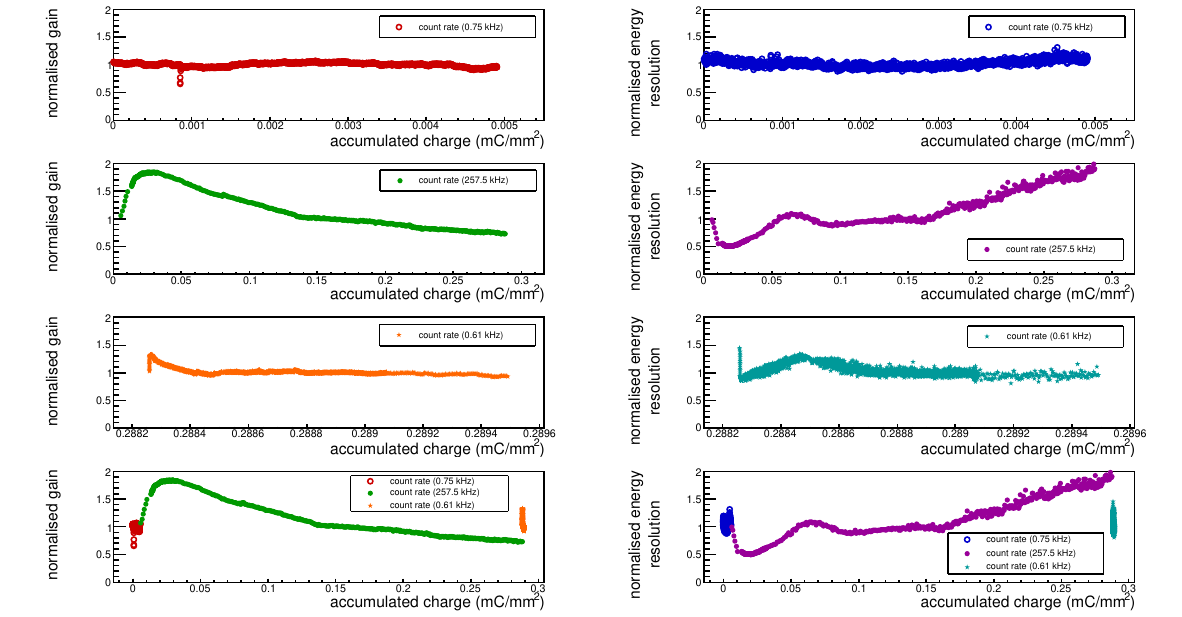}
\caption{(Colour online) Normalised gain and normalised energy resolution as a function of accumulated charge per unit area for the entire period and also for the three different rate phases.}
\label{fig6}
\end{figure*}

Figure~\ref{fig6} shows the normalised gain and normalised energy resolution as a function of the total charge accumulated per unit irradiated area of the GEM chamber, which is directly proportional to time. The accumulated charge per unit area ($\frac{dq}{dA}$) is calculated by

\begin{equation}
\centering
\frac{dq}{dA} = \frac{r \times n_{primary} \times e \times G \times dt}{dA}\label{chperarea}
\end{equation} 

where, $r$ is the measured rate in Hz incident on a particular area $dA$ of the detector, $n_{primary}$ (212 for Ar/CO$_2$ 70/30 ratio) is the number of primary electrons for a single X-ray photon, $e$ is the electronic charge, $G$ is the gain and $dt$ is the time in second. It is observed from Figure \ref{fig6} that the accumulated charge per unit area is much higher in the high-rate region compared to the low-rate phases, as expected from the larger irradiation flux. In the low-rate phases, both normalised gain and normalised energy resolution remain stable with respect to the accumulated charge per unit area, showing no significant signs of degradation. However, in the high-rate phase, the normalised gain first increases from the stabilised value of $\sim$~1 to a value $\sim$~2 and then stabilised again to $\sim$~1, reflecting the competing effects of charge accumulation and field modification inside the GEM holes. Again, when the irradiation returns from high rate to low rate, the normalised gain again increases and then stabilises, which is described as the phenomenon of reverse charging-up.

\begin{figure}[htb!]
\centering
\includegraphics[scale=0.4]{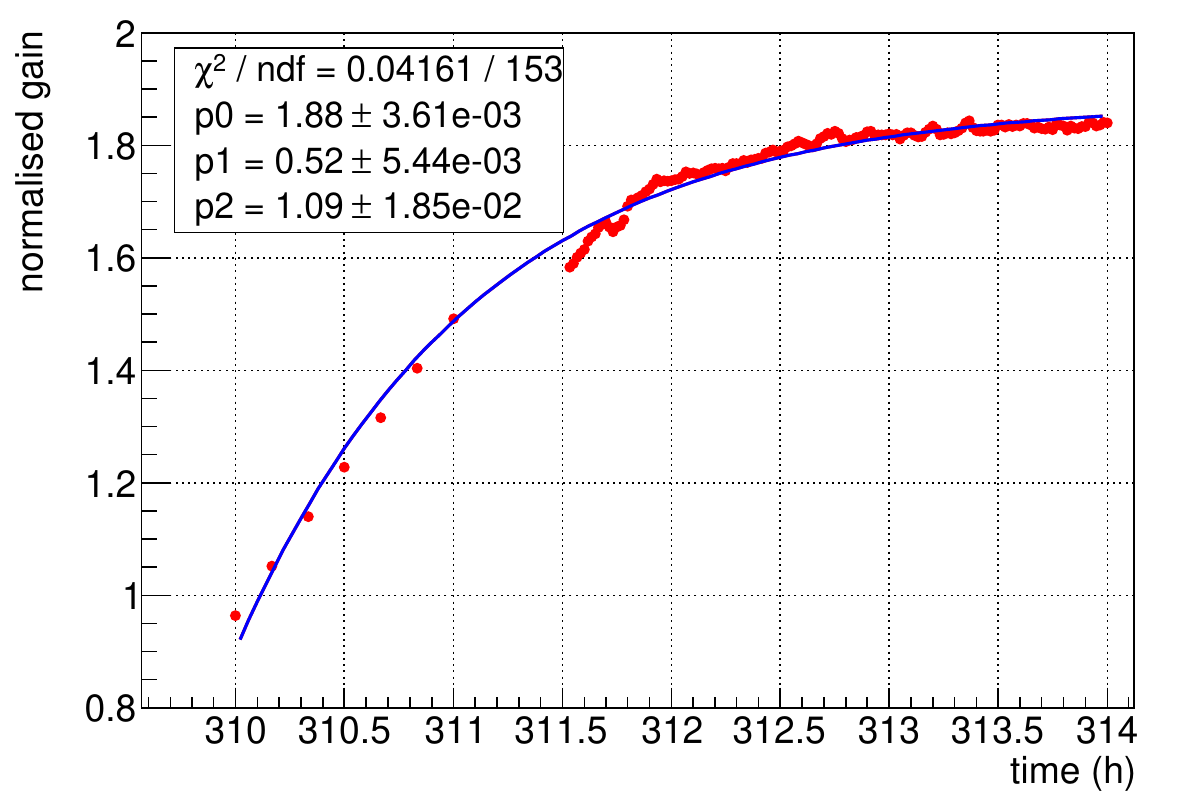}
\caption{(Colour online) Normalised gain as function of time (hour) for changing first low rates to high rates.}
\label{fig7}
\end{figure}
\begin{figure}[htb!]
\centering
\includegraphics[scale=0.4]{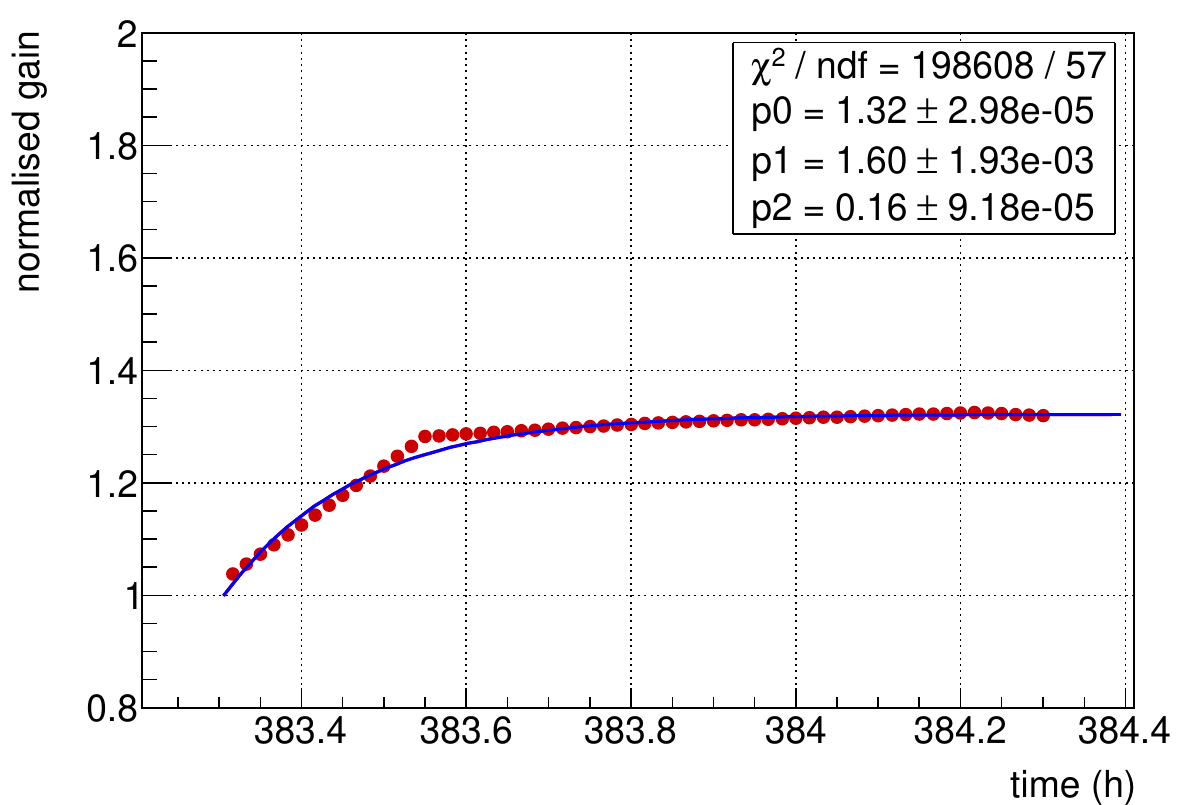}
\caption{(Colour online) Normalised gain as function of time (hour) for changing high rates to low rates.}
\label{fig8}
\end{figure}

The normalised gain as a function of time is zoomed for a period of about 4 hours from 310~h to 314~h to calculate the charging-up time and shown in Figure~\ref{fig7}. It is observed from Figure~\ref{fig7} that when the X-ray irradiation rate changes from low to high around 310 hours after the start of operation, the gain rises gradually before stabilisation, showing an exponential trend similar to the charging of an RC circuit. This charging-up behaviour is well described by the expression

\begin{equation}
G(t) = p_0 \left( 1 - p_1 e^{-t/p_2} \right) 
\end{equation}

where $G(t)$ is the instantaneous normalised gain, $p_0$, $p_1$ are constants and $p_2$ is the characteristic time constant, and $t$ is the elapsed time. This expressions provide a quantitative description of the charging-up phenomenon observed in the GEM detector operation~\cite{Silva2019,DelPapa2018}. In the present measurement, the time constant is found to be 1.09 hour for the low-to-high rate transition. This charging-up  phenomenon arises primarily from the accumulation of charges on the Kapton surfaces inside the GEM holes. As irradiation begins at a higher rate, electrons and ions created in avalanches drift towards the dielectric walls, where a fraction of them become trapped. The build-up of this surface charge modifies the local electric field in the hole: initially, the multiplication field is increased, causing the rise in the gain, until a balance is reached between the rate of charge deposition and the rate of charge leakage through the finite surface conductivity of the dielectric. Furthermore, at very high irradiation rates, ion space charge in the drift and transfer regions can temporarily alter the effective fields, leading to fluctuations superimposed on the exponential rise.

Overall, the charging-up effect reflects the complex interplay of surface charge accumulation, environmental influences, and space charge effects inside the GEM holes. These mechanisms together give rise to the observed RC-like exponential increase in gain until a steady operating state is established.

During the transition from high rate to low rate also the normalised gain as a function of time is zoomed and plotted in Figure~\ref{fig8}. It is observed from Figure~\ref{fig8} that when the particle rate decreases from a higher value to a lower value, the normalised gain of the GEM detector gradually increases. This behaviour is similar to the charging-up effect in GEM detectors. Since the increase in gain is observed in this case during the decrease of rate, it is defined as the reverse charging-up. In this case the characteristics time is found to be 0.16~h.

\section{Summary and Discussion}\label{sum}

In this work, the performance of a double-mask triple GEM detector is studied under continuous irradiation with an $^{55}$Fe X-ray source. The main observables are the gain, energy resolution, count rate, and their dependence on accumulated charge and ambient parameters.  

The measurements show that the detector operates stably during low irradiation phases, where both gain and energy resolution fluctuate only within small limits. Increase in the gain is observed because of charging-up of the dielectric medium inside the GEM holes during the transition from low rate exposure to a high rate exposure of X-ray. The energy resolution behaves inversely.

In presence of the high voltage, in this study when the rate of X-ray irradiation changed from high to low value, another increase in the normalised gain is observed. Since this is observed when the radiation rate is decreased, the phenomenon is defined as the reverse charging-up. In the case of reverse charging-up the characteristics time is found to be 0.16~h.

The charging-up behaviour of the gain follow exponential functions, closely resembling the response of an RC circuit. This simple analogy provides a useful way to describe and quantify the time constants of these processes. For the transition from low to high irradiation rate, the gain takes about 1.9~h to reach a stable value, while for the reverse transition from high to low rate the relaxation is slower, taking $\sim$~0.16~h.  

The take-home message from this work is that the increase in gain or charging-up is observed not only when the GEM detector is exposed to high radiation in presence of high voltage as dielectric Kapton is present inside the GEM foil, but also the increase in gain is observed when the rate of radiation is decreased from high to a low value. This kind of reverse charging-up phenomenon during the transition of radiation exposure from a high rate to low rate is reported for the first time.

Overall, these results shows that the double-mask triple GEM detector is capable of stable and reliable operation over long periods. The RC-type charging-up and reverse charging-up behaviour must be taken into account for precision measurements, but it does not limit the long-term performance of the detector. This makes the technology well suited for use in high-rate environments such as the CBM experiment at FAIR.

\section{Acknowledgements}
	
The authors would like to thank the RD51 collaboration (presently DRD1 collaboration) for the support in building and initial testing of the chamber at CERN. The authors would also like to thank Dr. Arindam Sen of Jagiellonian University, Poland, Dr. Sayak Chatterjee of  University of Massachusetts, Amherst, USA and  Dr. Somen Gope of Warsaw University of Technology, Poland for valuable discussions during the work. The authors acknowledges Mr. Subrata Das for helping in building the collimators used in this study. This work is partially supported by the research grant SR/MF/PS-01/2014-BI from DST, Govt. of India, and the research grant of the CBM-MuCh project from BI-IFCC, DST, Govt. of India. S. Mandal acknowledges his UGC-NET fellowship (NTA Ref. No.: 221610099585, Dt. 29.11.2022) for the support. S. Biswas acknowledges the support of the DST-SERB Ramanujan Fellowship (D.O.No. SR/S2/RJN-02/2012).


\end{document}